\documentclass[notitlepage,twocolumn,nofootinbib,10pt,showkeys,a4paper]{revtex4-1}

\usepackage{graphicx}
\usepackage{amsmath}
\usepackage{array}
\usepackage[colorlinks]{hyperref}

\newcommand{\tb}{t_\beta}

\newcommand{\mhl}{m_{h^0}}

\newcommand{\mhp}{m_{H^\pm}}

\newcommand{\fb}{\text{ fb}}

\newcommand{\gev}{\text{ GeV}}
\newcommand{\tev}{\text{ TeV}}

\newcommand{\sqrtHatS}{\sqrt{\hat s}}

\newcommand{\nlo}{\hat{\sigma}^\text{NLO}}
\newcommand{\lo}{\hat{\sigma}^\text{LO}}

\newcommand{\factor}{0.47}

\newcommand{\factorb}{0.41}

\usepackage{lineno}

\begin{document}

\title{Exploring the prominent channel:  Charged Higgs pair production in supersymmetric Two-parameter Non-Universal Higgs Model}

\author{Nasuf Sonmez}
\email{nasuf.sonmez@ege.edu.tr}
\affiliation{Department of Physics, Faculty of Science, Ege University, 35040 Izmir, Turkey}
\date{\today}


\begin{abstract}
In this study, the charged Higgs pair production is calculated in the context of the supersymmetry at a $\gamma\gamma$-collider. The channel is explored in Two-parameter Non-Universal Higgs Model where the model provides relatively light neutral and charged Higgs bosons. The computation is extended to one loop-level, and the divergence arising in the loop-diagrams are cured with the radiative photon correction. The production rate of the charged Higgs pair reaches up to $\hat{\sigma}_\text{UU}^\text{LO+NLO}=121\text{ fb}$ at $\sqrt{\hat{s}}=635\text{ GeV}$. The analysis of the cross-section is also given varying the parameters $m_A$ and $\tan\beta$. The total convoluted cross-section with the photon luminosity in an $e^+e^-$ machine is calculated as a function of the center-of-mass energy up to $1\text{ TeV}$, and it gets up to $42\text{ fb}$ at $\sqrt{s}=900 \text{ GeV}$ depending on the polarization of the initial electron and laser photon.
\end{abstract}
\pacs{11.30.Pb \sep 14.80.Ly \sep 14.80.Nb}
\maketitle


\section{Introduction}

In particle physics, interactions between the fundamental particles are expressed by symmetries. The invariance of these symmetries leads to the conservation of physical quantities and also imposes the equation of motion and the dynamics of the system. In this sense, the Standard Model (SM) is governed by local gauge symmetry, and it particularly has unitary symmetry $SU(3)\times SU(2)\times U(1)$. Besides local gauge symmetries, there is another one called Supersymmetry (Susy) \cite{Martin:1997ns}, which proposes a relationship between two basic classes of fundamental particles: bosons (particles with an integer-valued spin) and fermions (particles with half-integer spin). 
The Higgs boson was discovered at the LHC \cite{Aad:2012tfa, Chatrchyan:2012xdj, Aad:2015zhl}. Thus, the SM was completed. However, it is believed that the SM is a low energy approximation of a greater theory that defines new physics at higher scales. Susy is one candidate for these models which has attracted much attention.


The LHC running at $\sqrt{s}=7,8,13\tev$ energies made it possible to test many low scale Susy predictions. Unfortunately, no new particle has been discovered at the LHC.
That restricted the Susy parameter space \cite{Khachatryan:2014mma, Khachatryan:2014qwa, Khachatryan:2014doa, Altunkaynak:2015kia, Aaboud:2018zjf}; therefore, the masses of the sparticles were pushed to higher scales. It was concluded that the simple weak scale Susy picture was not valid, and thus, new scenarios were proposed in the Susy context. 
The Minimal Supersymmetric Standard Model (MSSM) \cite{Nilles:1983ge, Haber:1984rc} is an extension of the SM that realizes supersymmetry with a minimum number of new particles and new interactions. The simplified versions of the MSSM can be derived from a grand unified theory (GUT) under various assumptions. The model was generally defined with the following parameters: the soft-breaking masses $m_0$, $m_{1/2}$ and $A_0$ which are assumed universal at the GUT scale, the higgsino mixing mass parameter $\mu$, and the ratio of the two vacuum expectation values of the two Higgs doublets $\tan\beta$. 
Besides these parameters, the sparticles get a contribution from the so-called soft breaking terms. These terms miraculously give extra mass to these sparticles at the weak scale so that their masses are pushed to higher scales. These soft breaking terms are usually assumed universal at the GUT scale.
It is possible to assume non-universality for some of these parameters at the GUT scale with different motivations. For example: in \emph{one-parameter} Non-Universal Higgs Model (NUHM1), the soft-breaking contributions to the electroweak Higgs multiplets ($m^2_{H_d}$ and $m^2_{H_u}$) are equal, $m^2_{H_u}= m^2_{H_d}=sign(m_\phi)|m^2_\phi|$, but non-universal, $m^2_{H_u}(\text{GUT})=m^2_{H_d}(\text{GUT})\neq m_0^2$. 
Another one is called the \emph{two-parameter} Non-Universal Higgs Model (NUHM2) \cite{Ellis:2002iu, Ellis:2002wv}, and this time the soft-breaking contributions to the electroweak Higgs multiplets are not equal $m^2_{H_u}\neq m^2_{H_d}\neq m_0^2$. This scenario fits into all the current constraints on Susy, and most importantly, all electroweak observables get low contribution overall from radiative corrections. 
Thus, the NUHM2 scenario can be adjusted easily to get outside the limits obtained by the LHC. 
The masses of the squarks and the sleptons can be specially arranged above the TeV scale with the lightest electroweakinos ($\tilde{\chi}_1^{0,\pm}$) and the Higgses ($h^0, H^0, A^0, H^\pm$) being set deliberately below the TeV range. 
Among these particles, charged Higgs ($H^\pm$) boson emerges naturally in extended theories with at least two Higgs fields.
The observation of a single or pair production of charged Higgs boson is a hot research topic, and it has been studied extensively in all high-energy experiments. Its observation will be solid proof of an extended scalar sector.

The particle physics community has several proposals for future colliders to study the properties of the Higgs boson and investigate the extended scalar sector. In all these proposals, a lepton collider is planned; therefore, the next one will be an $e^+e^-$-collider. Additionally, an $e^+e^-$ collider can host a  $\gamma\gamma$-collider, basically high energy electrons are converted to high energy photons with a conversion rate of $k\approx 1$ by Compton back-scattering \cite{Telnov:1989sd} technique. 
The LHC has not found any trace about the extra Higgs bosons or the lightest electroweakinos. Therefore, one may wonder what is the potential of the future colliders about the supersymmetry searches, particularly, the charged Higgs pair production.
The charged Higgs pair in lepton colliders were studied extensively before in Refs. \cite{Beccaria:2002vd, Heinemeyer:2016wey}. Since the production rate in $e^+e^-$-collisions is s-channel suppressed, the cross-section of $\gamma\gamma \rightarrow H^+H^-$ is larger than that of $e^+e^-$-collider.
%
The production of the charged Higgs pair was investigated before at the next-to-leading order (NLO) for a $\gamma\gamma$-collider in Refs. \cite{Zhu:1997es, PhysRevD.53.1304} by taking into account only Yukawa and later full squark correction. However, the infrared divergence was not included there. The results showed that the one-loop corrections are from -20\% to +25\%. 
Later, the process was handled in Ref. \cite{Lei:2005kr} including the real radiative corrections. The calculation was presented for various benchmark points in MSSM, and the results showed that the correction is overall between -60\% to +5\% at $\sqrt{s}=1\tev$.
From the previous works which calculated the complete one-loop corrections, it is a proper question to ask what is the contribution of one-loop diagrams in the light of the exclusion limits set by the LHC and whether it is possible to observe the charged Higgs bosons in future colliders. Therefore, this process needs to be reevaluated.

In this study, the production of charged Higgs pair is studied, including one-loop corrections plus radiative ones. The cross-section of the charged Higgs pair is calculated as a function of the center-of-mass (c.m.) energy. The two polarization cases, $\hat{\sigma}_\text{RR}$ and $\hat{\sigma}_\text{RL}$, are calculated at the NLO level. Parameters of the model are varied, and the cross-section distributions are plotted. Additionally, the distributions are presented by convoluting the total cross-section with the photon luminosity in an $e^+e^-$-collider with various polarization configurations of the machine. 
The content of this paper is organized as follows:
In Sec. \ref{sec2}, the parameter space of the NUHM2 is defined.
In Sec. \ref{sec3}, one-loop Feynman diagrams, possible singularities in the calculation along with the procedure to cure these divergences are discussed thoroughly. In Sec. \ref{sec4}, numerical results of the total cross-section in the NUHM2 are delivered, and the conclusion is given in Sec. \ref{sec5}.



\section{Two-parameters non-universal Higgs model}
\label{sec2}

\begin{figure}[htbp]
\centering
\includegraphics[width=\factorb\textwidth]{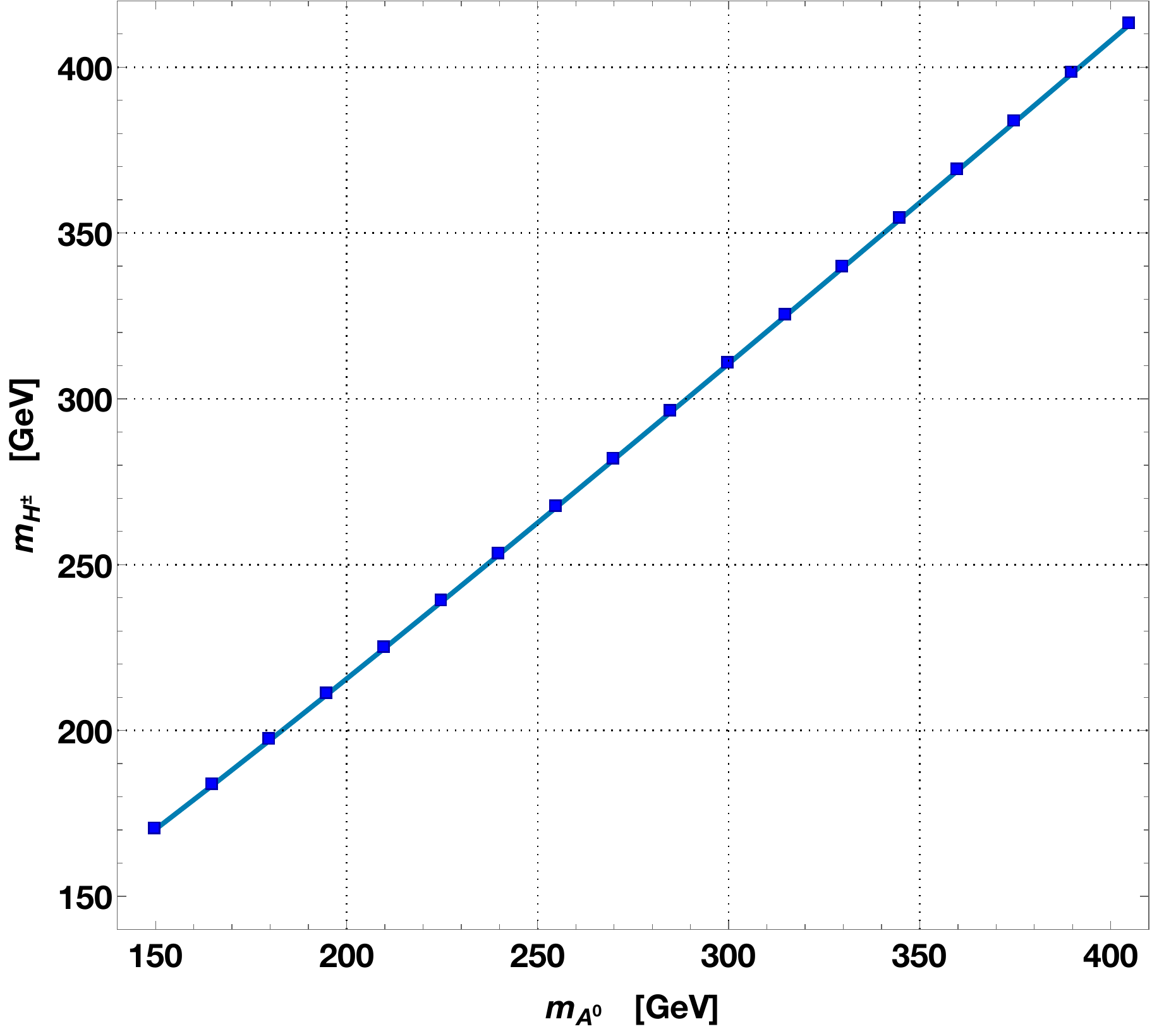}
\caption{\label{fig0}The charged Higgs mass ($m_{H^\pm}$) as a function of the input parameter $m_{A^0}$. }
\end{figure}

In Susy, the parameters at the GUT scale and the soft-breaking parameters are closely linked with the electroweak symmetry breaking. The results from the LHC pushed the previous limits on masses of the sparticles to high scales, and that resulted in the so-called "little hierarchy" between the weak scale and the scale of sparticles masses. The high scale fine-tuning of the large logarithmic contributions and the weak scale fine-tuning were needed to explain the little hierarchy.  The high scale tuning along with the weak scale tuning was discussed in detail in Ref. \cite{Baer:2013ula}. The NUHM2 is an effective model that is valid up to the scale $Q\leq M_\text{GUT}$, and the soft parameters at the GUT scale only serve to describe higher-dimensional operators of a more fundamental theory so that there might be correlations that cancel the large logs. Then, the implications on the weak scale become significant, and fine-tuning of the electroweak observables is needed. The electroweak fine-tuning parameter, which is defined in Ref. \cite{Baer:2012cf}, is given below:
\begin{equation}
        \Delta_\text{EW}=\text{max}_i |C_i|/(m_Z^2/2)
\end{equation}
where $C_{H_d}=m^2_{H_d^2}/(\tan\beta^2-1)$, $C_{H_u}=-m^2_{H_u^2}\tan\beta^2/(\tan\beta^2-1)$, and $C_{\mu}=-\mu^2$. The expression of the other two contributions at the one-loop level $C_{\Sigma_u^u(k)}$ and $C_{\Sigma_d^d(k)}$ were defined in Ref. \cite{Baer:2012cf}. Evidently, the $\Delta_\text{EW}$ gives the largest contribution to the mass of Z-boson. 
Since NUHM2 was inspired by GUT models where $\hat{H}_u$ and $\hat{H}_d$ belong to different multiplets, it is argued that the GUT scale masses $m^2_{{H}_u}$ and $m^2_{{H}_d}$ could be trade-off for the weak scale parameters $\mu$ and $m_A$ in Ref. \cite{Baer_2005}. To achieve small $\Delta_{EW}$, it is required that the soft mass parameter $|m^2_{H_u}|$, the mixing parameter of the Higgs-doublets $\mu$, and the radiative contribution $|\Sigma_u^u|$ have to be around $m^2_Z/2$ to within a factor of a few \cite{Baer:2012up, Baer:2012cf}. Then, the $|m_{H_u}^2|_\text{weak}$ and $\mu^2$ can have a value of $(100-300\gev)^2$ at any values of the parameters $m_0$ and $m_{1/2}$ in the non-universal Higgs models. Therefore $m_0=10\tev$ and $m_{1/2}=0.5\tev$ are assumed. The largest radiative correction, which is stop mixing, requires $A_0=\pm1.6\cdot m_0=\pm16\tev$, and finally $\mu=6\tev$ is assumed. Accordingly, the following two benchmark points with various ranges given in Table \ref{tab:param_table} are investigated in this paper. In Table \ref{tab:param_table}, all the sparticles are above the TeV except the Higgses and the elektroweakinos.
\begin{table}[htbp]
\caption{The input parameters in the NUHM2 scenario. All the values are in TeV. The mass spectrum and the mixing parameters are obtained using \textsc{IsaSugra} as a \textsc{SLHA} files.}
\begin{center}
\begin{tabular}{l|cccccccc}
    BP  &$m_0$  & $m_{1/2}$ &   $A_0$   & $t_\beta$ & $\mu$ &{$m_{A^0}$}    & $m_{H^\pm}$        \\ \hline
    1   &10     &    0.5    &   -16     & 7         & 6     &0.275          & 0.286              \\
    2   &10     &    0.5    &   -16     &(1-50)     & 6     &(0.15-0.4)     & (0.170-0.413)
\end{tabular}
\end{center} \label{tab:param_table}
\end{table}
The benchmark point 1 (BP-1) is taken from Ref. \cite{Baer:2013ula} \footnote{The SLHA files for this benchmark point were obtained from \url{http://flc.desy.de/ilcphysics/research/susy}.}. 
The range of  $\mhp$ in BP-1 ($170 < m_{H^\pm} < 413 \gev$) might be light considering the constraints from the B physics. It should be noted that the value of $BR(b \rightarrow s\gamma)= 4.6 \times 10^{â4}$ (given in \cite{Baer:2013ula}) in the BP-1 is excluded in context of Type-II Two Higgs doublet model \cite{Haller:2018nnx}. Accordingly, 
 The mass of charged Higgs boson being less than 600 GeV is not consistent with experimental data of $b \rightarrow s\gamma$, of course, this does not directly apply to the MSSM. 
The masses of all the Higgses and the elektroweakinos are less than $500\gev$, but the rest of the sparticles are still beyond the TeV, so they are also beyond the reach of the LHC. In BP-2, the parameters $m_A$ and $\tan\beta$ are varied in the following ranges $(1-50)$ and $(150-400)\gev$, respectively. The sparticle's masses and their mixing parameters are calculated with the help of \textsc{IsaSugra-v7.88}  \cite{Paige:2003mg}. 
Then one may wonder how the masses of the charged ($\mhp$) and the lightest Higgs bosons ($\mhl$) depend on the input parameter $m_{A^0}$.  In Fig. \ref{fig0}, the charged Higgs mass is plotted as a function of $m_{A^0}$, and it can be seen that the higher-order corrections (implemented in \textsc{IsaSugra}) lower the $\mhp$ compared to the tree-level mass relation $m_{H^\pm}^2=m_{W^\pm}^2+m_{A^0}^2$.
Besides, the \textsc{IsaSugra} calculated the lightest Higgs boson ($m_{h^0}$) mass between $120\leq m_{h^0}\leq125\gev$ in all the points defined in Table \ref{tab:param_table}. Therefore, $ m_{h^0}$ is consistent with 
the discovered Higgs boson at the LHC \cite{Aad:2015zhl}.

\section{The calculation at the Next-to-leading order}
\label{sec3}
In this section, the analytical expressions related to the cross-section of the charged Higgs boson pair
are provided. The scattering process is denoted as follows:
    \begin{eqnarray*}
    \gamma (k_1,\mu)+\;\gamma(k_2,\nu)\;\rightarrow \;  H^+ (k_3)+\; H^-(k_4) \; (i, j=1,2)\,,\nonumber
    \end{eqnarray*}
where $k_a$ $(a=1,...,4)$ are the four momenta of the incoming photons and outgoing charged Higgses, the parameters $\mu$ and $\nu$ represent the polarization vectors of the incoming photons.


\subsection{The process at the tree-level}

\begin{figure}[htbp]
\centering
\includegraphics[width=\factor\textwidth]{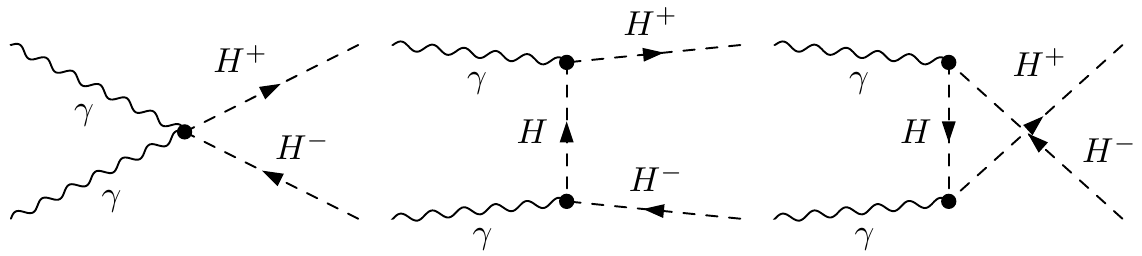}
\caption{\label{fig1}    Tree level Feynman diagrams for the charged Higgs pair production via photon-photon fusion.}
\end{figure}

The Feynman diagrams, which take place in the charged Higgs pair production via $\gamma \gamma$ collision at the tree-level, are plotted in Fig. \ref{fig1}. These diagrams and the corresponding amplitudes are constructed using \textsc{FeynArts} \cite{Hahn:2000kx}. The vertices were defined in the model file where all the couplings follow the convention given in Ref. \cite{Haber:1984rc}, and the \textsc{FeynArts} implementation of these rules were given in Ref. \cite{Hahn:2001rv}. After obtaining the total amplitude for the process, further evaluation is employed in \textsc{FormCalc} \cite{Hahn:2000jm, Hahn:2010zi}. 
The summation over the helicities of the final states, and averaging over the polarization vectors are determined by the \textsc{FormCalc} automatically. The cross-section of the polarized $\gamma\gamma$ collision at the tree level is defined as follows:        \begin{equation}
        \hat{\sigma}^{\text{LO}}(\hat{s};\mu,\nu)=\frac{ \lambda( \hat{s},m_{H^\pm}^2 )}{16 \pi \hat{s}^2} \left(\frac{1}{4} \sum_{hel}{|\mathcal{M}_\text{tree}(\mu,\nu)|^2}\right)\,,
        \label{eq:partcross}
        \end{equation}
where $\lambda( \hat{s},m_{H^\pm}^2 )=\sqrt{ \hat{s}^2/4-\hat{s}\,m^2_{H^\pm} }$, the parameters $\mu$ and $\nu$ define the polarization of the photons, $\hat{s}$ represents the c.m. energy in $\gamma\gamma$ collision, and the factor $1/4$ is the average of photon's polarizations. 
The Feynman diagrams are given in Fig. \ref{fig1} show that the couplings $g_{\gamma\gamma H^+H^-}$ and  $g_{\gamma H^+H^-}$ play a role at the tree-level cross-section. The couplings of a photon with charged particles are universal. Therefore, this process is a QED process at the tree level, and the cross-section depends on the mass of the charged Higgs. At the loop-level, the total cross-section could get significant contributions from the sparticles, and that requires a detailed analysis.
       
\begin{figure}[htbp]
\centering
\includegraphics[width=\factor\textwidth]{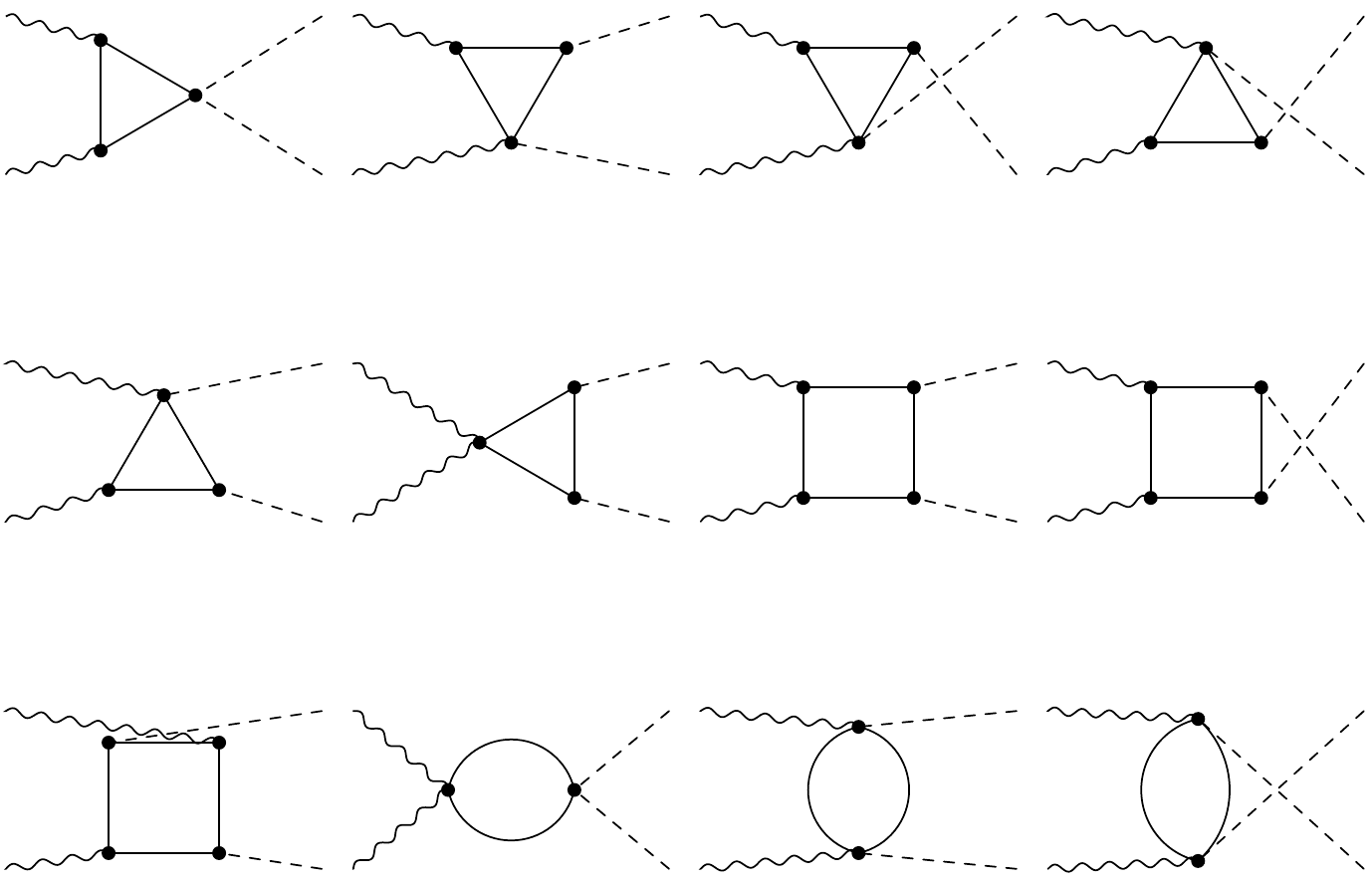}
\caption{\label{fig2} 
One-loop Feynman box diagrams for the charged Higgs pair production via photon-photon fusion. The wavy lines represent the incoming photons, dashed lines are the charged Higgses at the final state, the intermediated lines represent all the particles and the sparticles in the theory.}
\end{figure}
\begin{figure}[htbp]
\centering    
\includegraphics[width=\factor\textwidth]{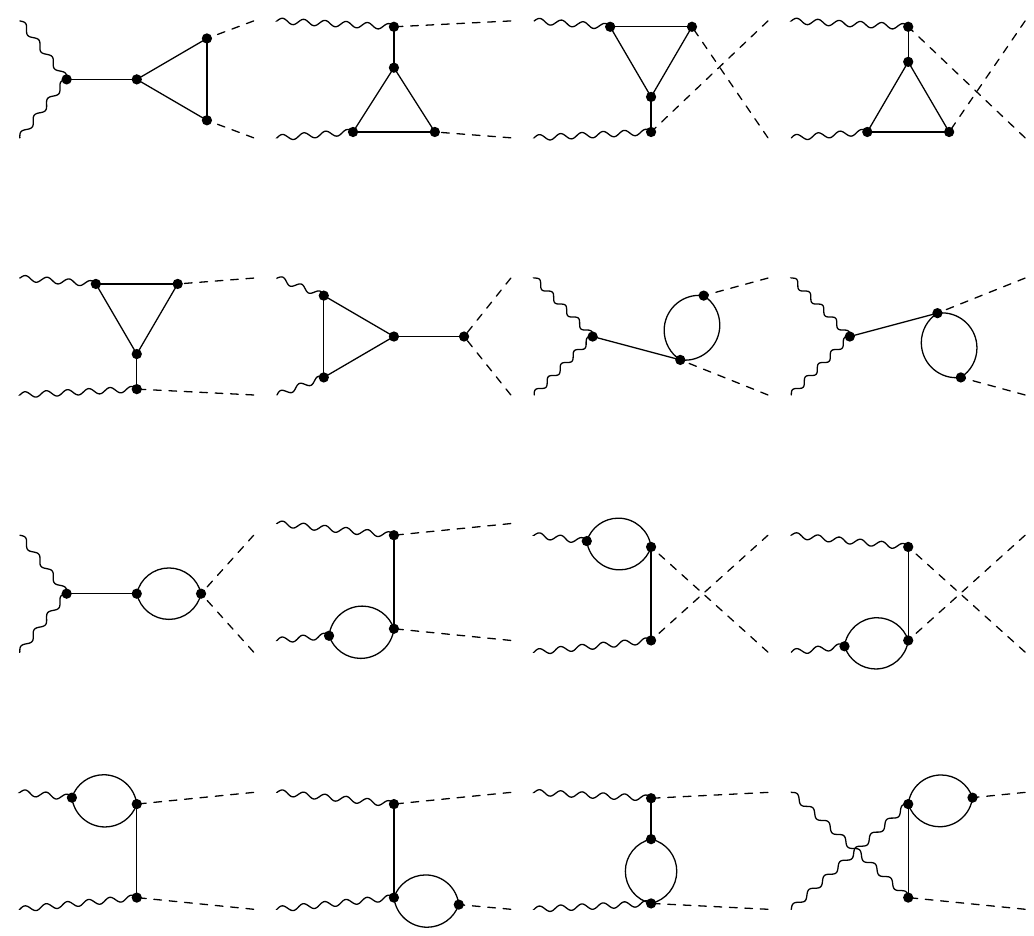}
\caption{\label{fig3} Triangle- and bubble-type s-channel Feynman diagrams for charged Higgs pair via photon-photon fusion. The wavy lines represent the incoming photons, dashed lines are the charged Higgses at the final state, the intermediated lines represent all the particles and the sparticles in the theory.}
\end{figure}
\begin{figure}[htbp]
\centering    
\includegraphics[width=\factorb\textwidth]{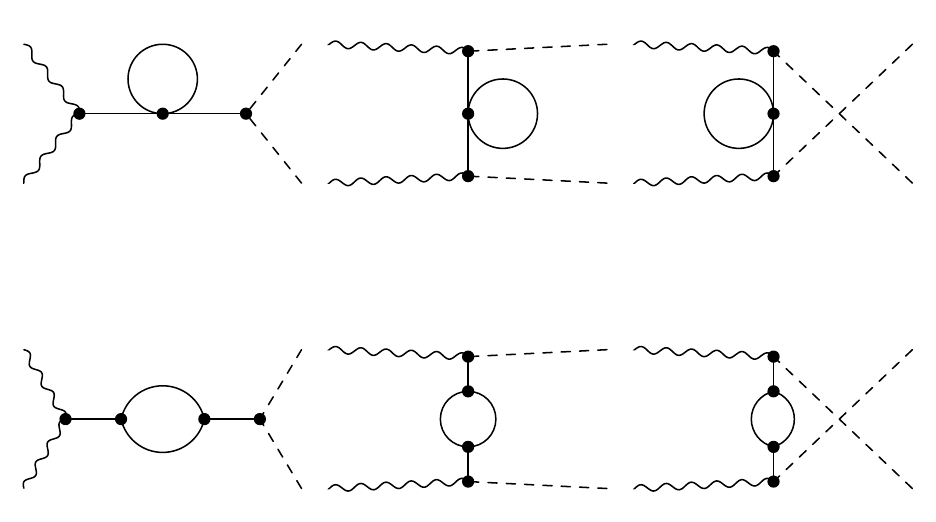}
\caption{Feynman self-energy diagrams for the charged Higgs pair production via photon-photon fusion. The wavy lines represent the incoming photons, dashed lines are the charged Higgses at the final state, the intermediated lines represent all the particles and the sparticles in the theory.}
    \label{fig4}
\end{figure}

The one loop-level calculation followed next.
Then the one loop-level diagrams are obtained with the help of FEYNARTS and grouped into three distinct topological sets of diagrams: self-energy diagrams, triangle- and bubble-type s-channel diagrams, and finally box-type diagrams. All these diagrams are given in  Fig. \ref{fig2}-\ref{fig4} where the intermediated lines between the initial and the final states represent the propagators of all the possible SM and Susy particles.


The corresponding total Lorentz invariant matrix element for the process at the one-loop level can be written as a sum over all these three contributions: the box-diagrams (Fig. \ref{fig2}), the triangle- and bubble-type diagrams (Fig. \ref{fig3}), and the self-energy diagrams (Fig. \ref{fig4}).
    \begin{equation}\label{eq6}
    \cal{M}_\text{virt}=\cal{M}_{\text{box}}+ \cal{M}_{\text{tri}}+ \cal{M}_{\text{self}} 
    \end{equation} 
    The one-loop virtual correction is calculated by the following formula where the squared term $|{\cal M}_\text{virt}|^2$ is not included to the calculation because it is very small.
\begin{equation}
\hat{\sigma}_{\text{virt}}(\hat{s})=\frac{ \lambda( \hat{s},m_{H^\pm}^2 )}{16 \pi \hat{s}^2} \frac{1}{4} \sum_{hel} 2 \text{Re} \left[ {\mathcal{M}^*_\text{tree}\mathcal{M}_\text{virt}} \right]\,
\label{eq:virtcross}
\end{equation}
where $\hat{s}$ represents the c.m. energy in the  $\gamma\gamma$ collision frame, and $\lambda( \hat{s},m_{H^\pm}^2 )$ is the same function defined before.


\subsection{Ultraviolet and infrared divergences}

\begin{figure}[htbp]
        \centering
            \includegraphics[width=\factorb\textwidth]{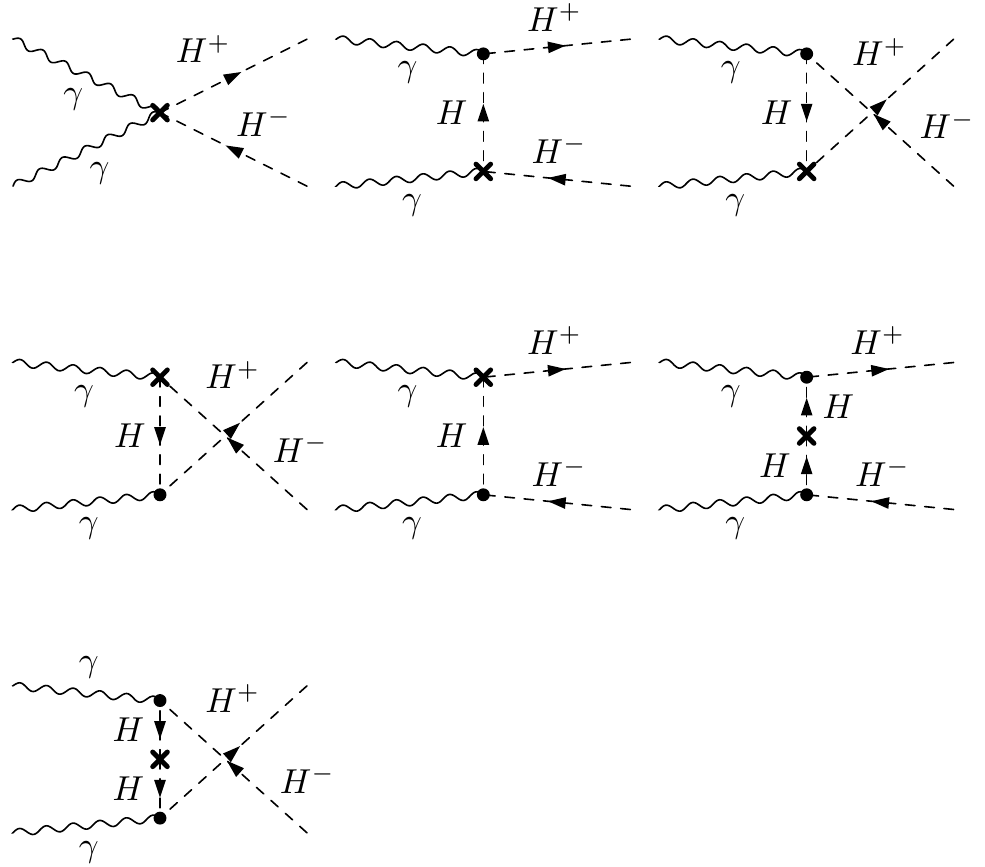}
        \caption{The Feynman diagrams that regularize the divergent vertices in the charged Higgs pair production.}
        \label{fig5}
        \end{figure}

In multi-loop calculations, an ultraviolet (UV) divergence occurs due to the contribution of terms with unbounded energy, or because of investigating the physical phenomena at an infinitesimally small distance. Since infinity is unphysical, the ultraviolet divergences require special treatment, and they are removed by the regularization procedure, which is called the renormalization. 
The divergent integrals are cured by including the counterterms, which simply regularize the divergent vertices; thus, the result becomes finite.
The renormalized MSSM model file in \textsc{FeynArts} follows the conventions of Refs. \cite{Haber:1984rc, Gunion:1984yn, Gunion:1989we}. The calculation and implementation of counterterms are described in Ref. \cite{Fritzsche:2013fta}.
The numerical calculation is performed in the \emph{'t Hooft-Feynman} gauge so that the propagators have a simple form, and the calculation of the loop integrals consumes less computing power. The constrained differential renormalization \cite{Siegel:1979wq} is employed because it is equivalent to the dimensional reduction \cite{delAguila:1998nd, Capper:1979ns} at the one-loop level \cite{Hahn:1998yk}. 
The cross signs in Fig. \ref{fig5} indicate all the vertices that require renormalization. They are included in the calculation, therefore, the total amplitude ${\cal M}_\text{virt}$ becomes UV finite.
The scalar and the tensor one-loop integrals are computed with the help of \textsc{LoopTools} \cite{vanOldenborgh:1989wn, Hahn:2000jm, Hahn:2010zi}, and the UV divergence is tested by changing the parameters $\mu$ and $\Delta$ on a large scale. The cross-section is stable in the numerical precision. That proves the divergence is removed in the calculation.

\begin{figure}[htbp]
\centering
\includegraphics[width=\factor\textwidth]{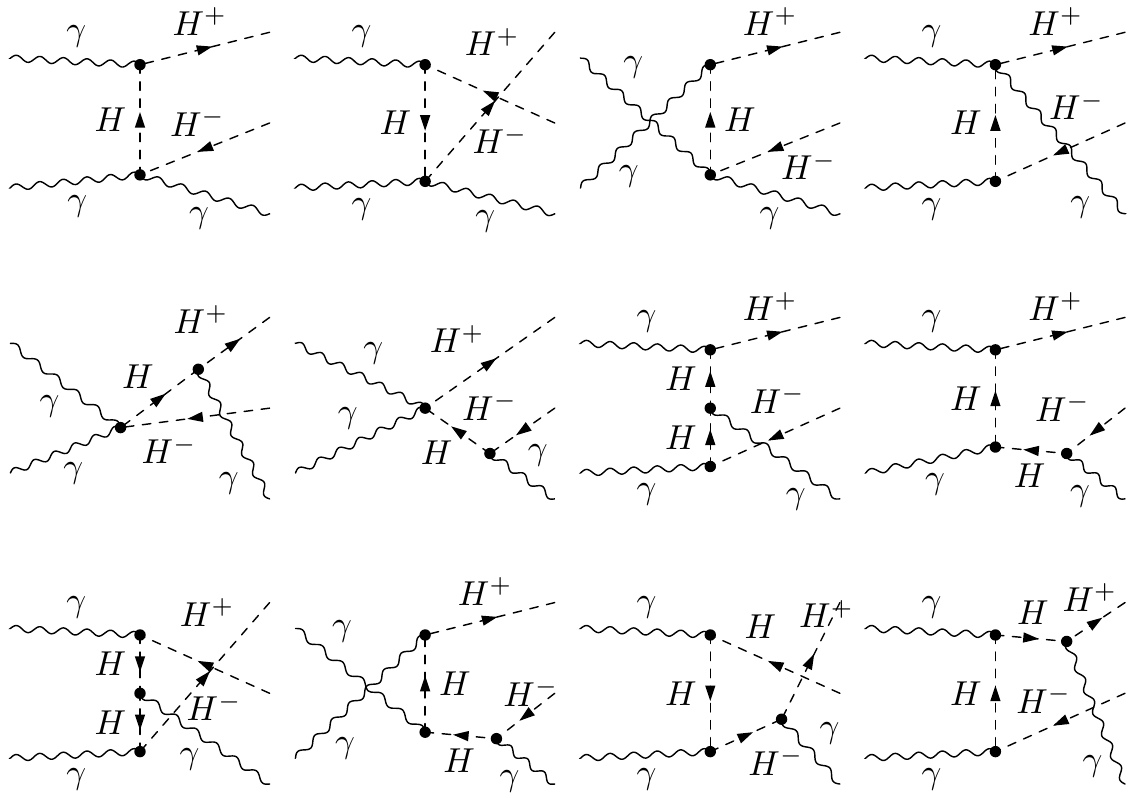}
\caption{The Feynman diagrams that contribute to the radiative corrections in charged Higgs pair production. The process $\gamma\gamma\rightarrow H^{+}H^{-}\gamma$ is plotted.}
        \label{fig6}
\end{figure}

In the computation, another type of singularity occurs due to the charged particles at the final state and the massless particles which are propagating with very small energy in the loops. 
This kind of divergence is called infrared divergence (IR), and if the photon had a mass of $\lambda$, then these divergent terms would be proportional to $\log \lambda$. According to the Kinoshita-Lee-Nauenberg theorem, these logs cancel in sufficiently inclusive observables. However, in non-confined theories such as the SM, it is possible to obtain substantial effect due to small fermion masses in non-inclusive observables. These IR divergent contributions are canceled by including similar divergences coming from phase-space integrals of the same process with additional photon radiation at the final state. In other words, a measurement acquired in an experiment intrinsically possesses this interaction already. The apparatus measures with a minimum energy threshold, and therefore, it could not measure the photons that might have been emitted with an energy of less than $\Delta E$. The cross-section of emitting soft photons at the final state has the same kind of singularity with massless photons propagating in the loops, and adding these contributions cancels the IR divergences in the calculation. The diagrams having an additional photon at the final state are given in Fig. \ref{fig6}.

\begin{figure}[htbp]
 \centering      
 \includegraphics[width=\factor\textwidth]{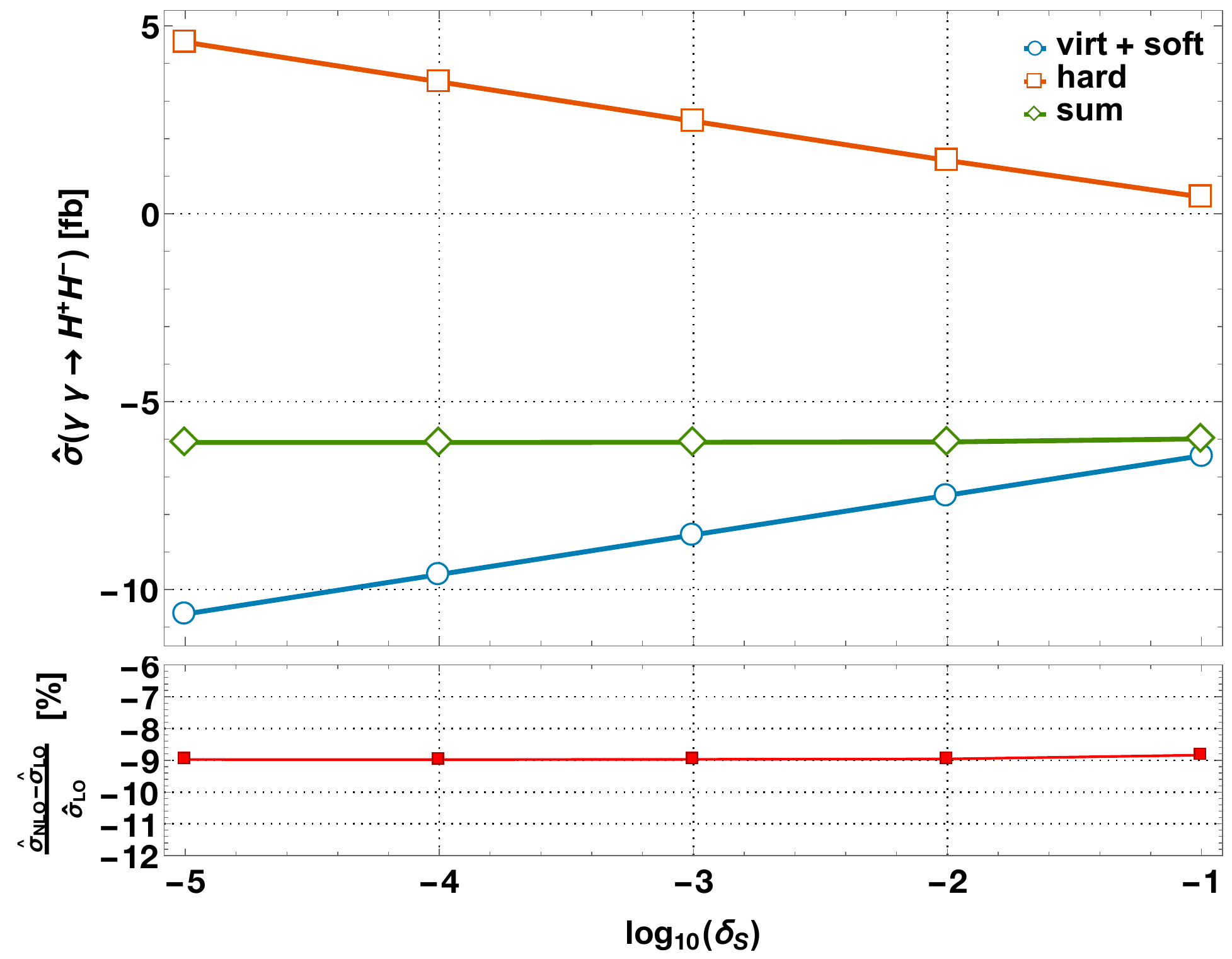}
 \caption{\label{fig7} All the corrections are plotted as a function of $\delta_s = \Delta E/\sqrt{\hat{s}}/2$. The orange line with the square marks represents the hard photon radiation, the light blue line with the circle marks line represents the virt+soft contribution, and the green line with the diamond marks shows the sum of all the contributions. The calculation is performed at $\sqrt{\hat{s}}=1.0 \tev$, the BP-1 is employed, and $\lambda=1$ is set.}
\end{figure}

The soft photon radiation correction is implemented in \textsc{FormCalc} following the description given in Ref. \cite{Denner:1991kt}. Overall, the correction is proportional with the tree level process; $\left( \frac{d\sigma}{d\Omega}\right)_\text{s}=\delta_\text{SB} \left( \frac{d\sigma}{d\Omega}\right)_\text{born} $, where $\delta_\text{SB}$ is the soft bremsstrahlung factor, and its explicit form is given in Ref. \cite{Denner:1991kt}. The factor $\delta_\text{SB}$ is a function of $\Delta E$, and $\Delta E$ separates the soft and the hard photon radiation. The photons are considered soft, if their energy is less than $\Delta E =\delta_s E=\delta_s \sqrt{\hat{s}}/2$. Summing this contribution  ($\sigma_{\text{soft}}(\lambda,\Delta E)$) with the virtual ($\sigma_{\text{virt}}(\lambda)$) one drops out the dependence on the photon mass parameter $\lambda$. However, the result now depends on the detector dependant parameter $\Delta E$, and the contribution coming from the hard photon radiation needs to be added as well to drop that dependence out. Thus a complete picture of the process is obtained. In conclusion, the total one-loop corrections could be written as a sum of the virtual, the soft photon radiation, and the hard photon radiation
\begin{eqnarray}
            \hat{\sigma}^{\text{NLO}}&=&\hat{\sigma}_{\text{virt}}(\lambda)+\hat{\sigma}_{\text{soft}}(\lambda,\Delta E)+\hat{\sigma}_{\text{hard}}(\Delta E).
\end{eqnarray}
Next, the same divergent test is employed on the parameter $\lambda$; it is varied on a large scale, and the sum of the virtual and the soft photon radiation becomes stable. Finally, the total cross-section needs to be checked for the detector dependant parameter, and the $\delta_s$ is varied logarithmically in Fig. \ref{fig7}. The virtual + the soft correction is plotted by a blue line, the hard photon radiation is plotted by orange, and the sum of all are indicated by a straight green line. The calculation is done at $\sqrt{\hat{s}}=1\tev$ with the parameters defined in benchmark point 1 in Table \ref{tab:param_table}. The total NLO correction is around $\sim -9\%$ compared to the LO cross-section. The straight green line in Fig. \ref{fig7} indicates that the sum of all the contributions is stable in changing the detector dependent parameter $\delta_s$ on a large scale. 


\subsection{Convoluting the cross-section with the photon luminosities}
    

In real life, getting a high intensity monochromatic photon beam is technically difficult and also might be costly. Instead, an $e^+e^-$-collider can be used to extract a photon beam by Compton back-scattering technique. It is argued that the big fraction of the c.m. energy of the electron beam can be transferred to the Compton back-scattered photons in Ref. \cite{Ginzburg_1984}. 
%
Then, the scattering process looks as  $e^+e^-\rightarrow(\gamma\gamma)\rightarrow H^+ H^-$, and the production rate can be obtained by convoluting the partonic cross-section $\hat{\sigma}_{\gamma\gamma\rightarrow H^+H^-}(\hat{s})$ with the photon luminosity in an $e^+e^-$-collider. The convolution is defined as follows:
\begin{eqnarray}
\label{eq:total_cross}
\sigma(s)=\frac{1}{4} \sum_{\alpha,\beta=\pm1} \int_{4m_{H^\pm}^2/s}^{y_m^2} d\tau  &&\frac{dL_{\gamma\gamma}}{d\tau} \left(1+\alpha\xi(y)\right)\\ \nonumber
 &&\left(1+\beta\xi(\tau/y)\right)\hat{\sigma}_{\alpha\beta} ,
\end{eqnarray}   
where $\hat{\sigma}_{\alpha\beta}=\hat{\sigma}^{\text{LO}}+\hat{\sigma}^{\text{NLO}}=\hat{\sigma}_{\gamma^\alpha\gamma^\beta\rightarrow  H^+H^-}( \hat{s})$ is the scattering cross-section for the polarization configurations of the incoming photon beams, $(\alpha, \beta)$ represent the polarization of the photons. The $s$ and the $\hat{s}$ are the c.m. energy in $e^+e^-$ collisions and $\gamma\gamma$ sub-process, respectively therefore $\tau=\hat{s}/s$ represents ratio of the c.m. energy in $e^+e^-$ collisions and $\gamma\gamma$ sub-process. 
The photon luminosity is defined as follows:
\begin{equation}
\frac{dL_{\gamma\gamma}}{d\tau}=\int_{\tau/y_{m}}^{y_{m}}\frac{dy}{y}F_{\gamma}(x,y)F_{\gamma}\left(x,\tau/y\right)\,.
\end{equation}
The energy spectrum $F_{\gamma}(x,y)$ and the mean polarization $\xi(y)$ of the scattered photons were defined in Refs. \cite{Ginzburg:1982bs, Ginzburg_1984, Telnov:1989sd, Telnov:1995hc}, where $y=E_\gamma/E_e$ with $E_\gamma$ and $E_e$ being the energy of photon and electron beams, respectively. 
In this study, the energy spectrum of the photons includes only the Compton back-scattered photons, and nonlinear effects are not taken into account. The maximum fraction of the photon energy is defined as $y_m=x/(1+x)$ where $x=\left(4 E_e  E_l /m_e^2\right)$, the laser photon energy $E_l=1.17\text{ eV}$, $m_e$ is the electron mass, and we set $x=4.8$ in the calculation \cite{Ginzburg_1984, Telnov:1995hc}.


\section{Numerical Results and Discussion}
\label{sec4}
    
The following input parameters in the SM were given in Ref. \cite{Eidelman:2004wy} where $m_W = 80.399\gev$, $m_Z=91.1887\gev$, $m_t=173.21\gev$, and $\alpha(m_Z)=1/127.944$. The benchmark points are defined in Tab. \ref{tab:param_table}, and the following results are obtained using the \textsc{SLHA} files.
    
\begin{figure}[htbp]
\centering
\includegraphics[width=\factor\textwidth]{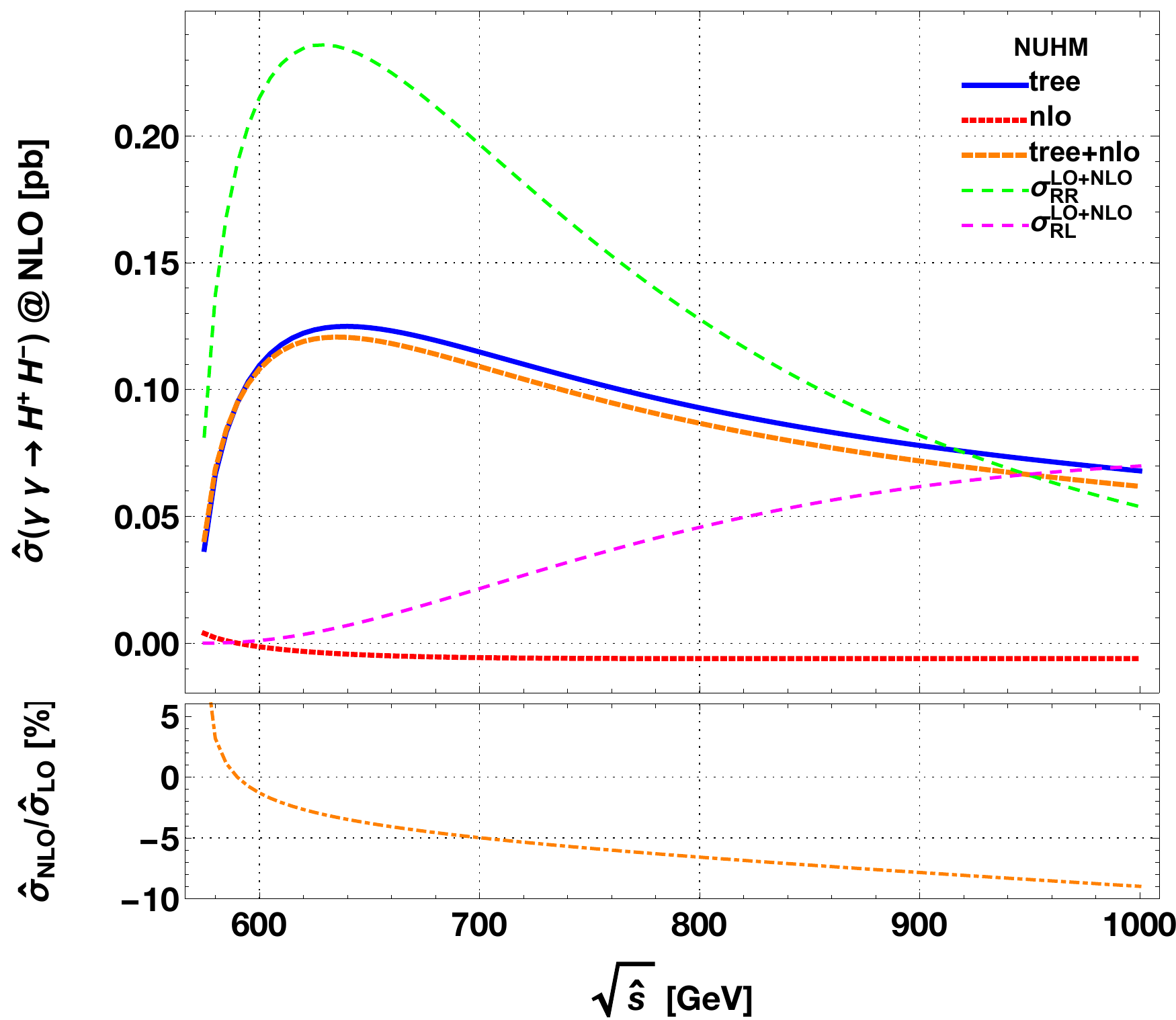}     
\caption{\label{fig8}The cross-section distributions of the process $\gamma\gamma\rightarrow H^+ H^-$ and the ratio $\hat{\sigma}_\text{NLO}/\hat{\sigma}_\text{LO}$ are plotted as a function of $\hat{s}$ in NUHM2 scenario. Two polarization cases $\lo_\text{RR}+\nlo_\text{RR}$ and $\lo_\text{RL}+\nlo_\text{RL}$ are indicated by green-dotted and magenta-dotted lines in the figure, respectively.}
  \end{figure}

The distribution of the tree level, the loop-level, and their sum are plotted in Fig. \ref{fig8} as a function of c.m. energy up to $\sqrtHatS=1\tev$; the ratio $\nlo/\lo$ is added at the bottom. The partonic cross-section rises sharply when the c.m. energy passes the total mass of the charged Higgs pair, which is 580 GeV here, and the unpolarized cross-section reaches up to $125\fb$ at $\sqrtHatS=0.64\tev$. The total corrected cross-section starts falling moderately after $\sqrtHatS=0.64\tev$, and it decreases to $62\fb$ at $\sqrtHatS=1\tev$. The distributions of two polarization cases, the $\hat{\sigma}_\text{RR}$ ($J_z=0$) and the $\hat{\sigma}_\text{RL}$ ($J_z=2$), are also given in Fig. \ref{fig8} by green-dashed and magenta-dashed lines, respectively. It shows that the total cross-section is dominated by $J_z=0$ for $\sqrtHatS\lesssim 0.95\tev$, and $J_z=2$ gets higher in $\sqrtHatS>0.95\tev$. Overall, the sum of the virtual and the real corrections (NLO) is negative for $\sqrtHatS>0.59\tev$, and thus it lowers the tree-level cross-section by $\sim 9\%$. Since the masses of the sparticles are beyond the TeV, their contributions to the cross-section via loops are small, but they still have an impact.

\begin{figure}[htbp]
\centering
\includegraphics[width=\factor\textwidth]{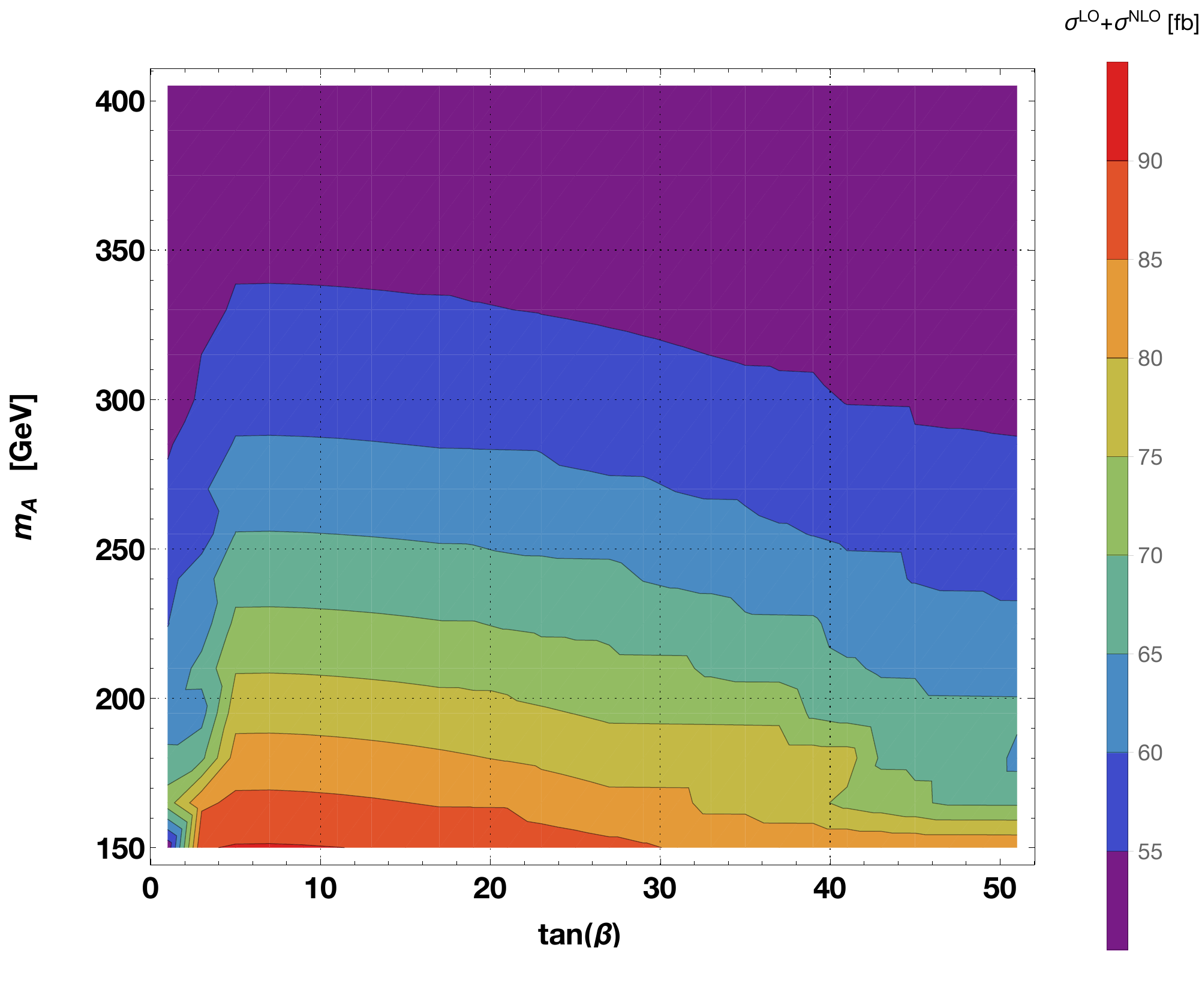}     
\includegraphics[width=\factor\textwidth]{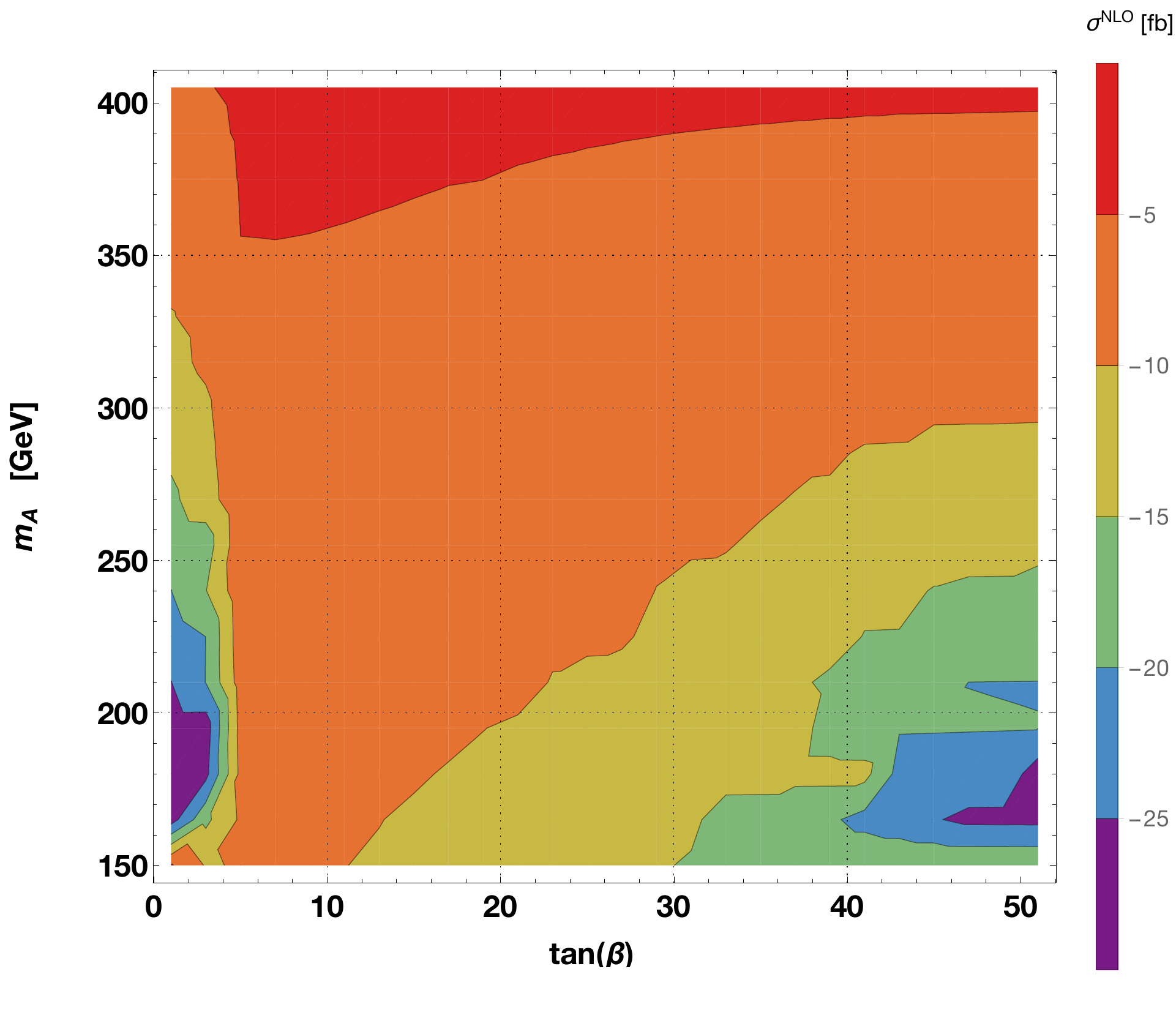}     
\includegraphics[width=\factor\textwidth]{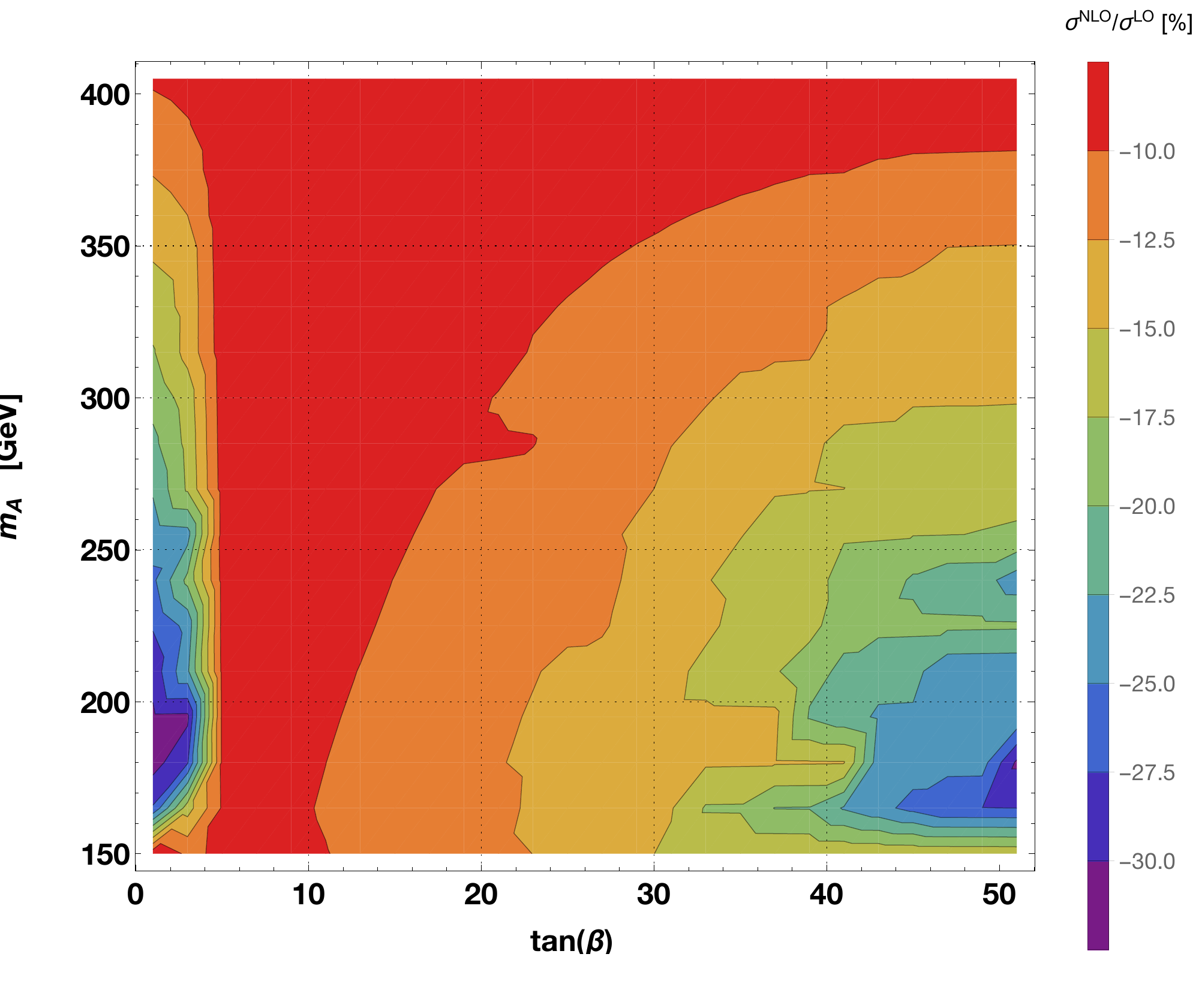}     
\caption{\label{fig10}
The cross-section distributions of the process $\gamma\gamma\rightarrow H^+ H^-$ with the unpolarized incoming photons are plotted at $\hat{s}=1\tev$. 
(top): $\lo+\nlo$, 
(middle): $\nlo$,
(bottom): the ratio $\nlo/\lo$ is plotted as a function of $\tan\beta$ and $m_A$.}
\end{figure}

In Fig. \ref{fig10}, various cross-section distributions are plotted as a function of $\tan\beta$ and $m_A$ for the region defined in BP-2 at $\sqrtHatS=1\tev$.
At the top, the total cross-section distribution ($\lo+\nlo$) is given. As it is expected, the cross-section is higher at low $m_A$ values because the $m_A$ is directly related to the charged Higgs boson, so the phase space of the final state particles. The cross-section gets smaller with increasing $\tan\beta$, but the change is low. 
Since the cross-section at the tree-level is mostly QED, all the model related contribution comes from the loop-level. Therefore, the one-loop level contribution, $\nlo$, is given in Fig. \ref{fig10} (middle) as a function of $\tan\beta$ and $m_A$. The $\nlo$ is negative in the whole region, and it decreases the tree-level cross-section up to $25\fb$, but it is around $\sim-10\fb$ for most of the parameter space.
In Fig. \ref{fig10} (bottom), the ratio  $\nlo/\lo$ is plotted, and it shows that the total one-loop and soft photon corrections could decrease the LO cross-section by $30\%$ at most. The ratio is always negative in the whole region. It gets lower at large $\tan\beta$ and low $m_A$ values plotted by dark blue regions. The ratio is the highest in the  $4\lesssim\tan\beta\lesssim 10$ region, and it expands at high $m_{A^0}$ values. The total cross-section is lowered at $\tan\beta \lesssim 4$. Overall, the figure shows that the one-loop contribution is negative, and it lowers the cross-section at $\sqrt{s}=1\tev$.

\begin{figure}[htbp]
\centering
\includegraphics[width=\factor\textwidth]{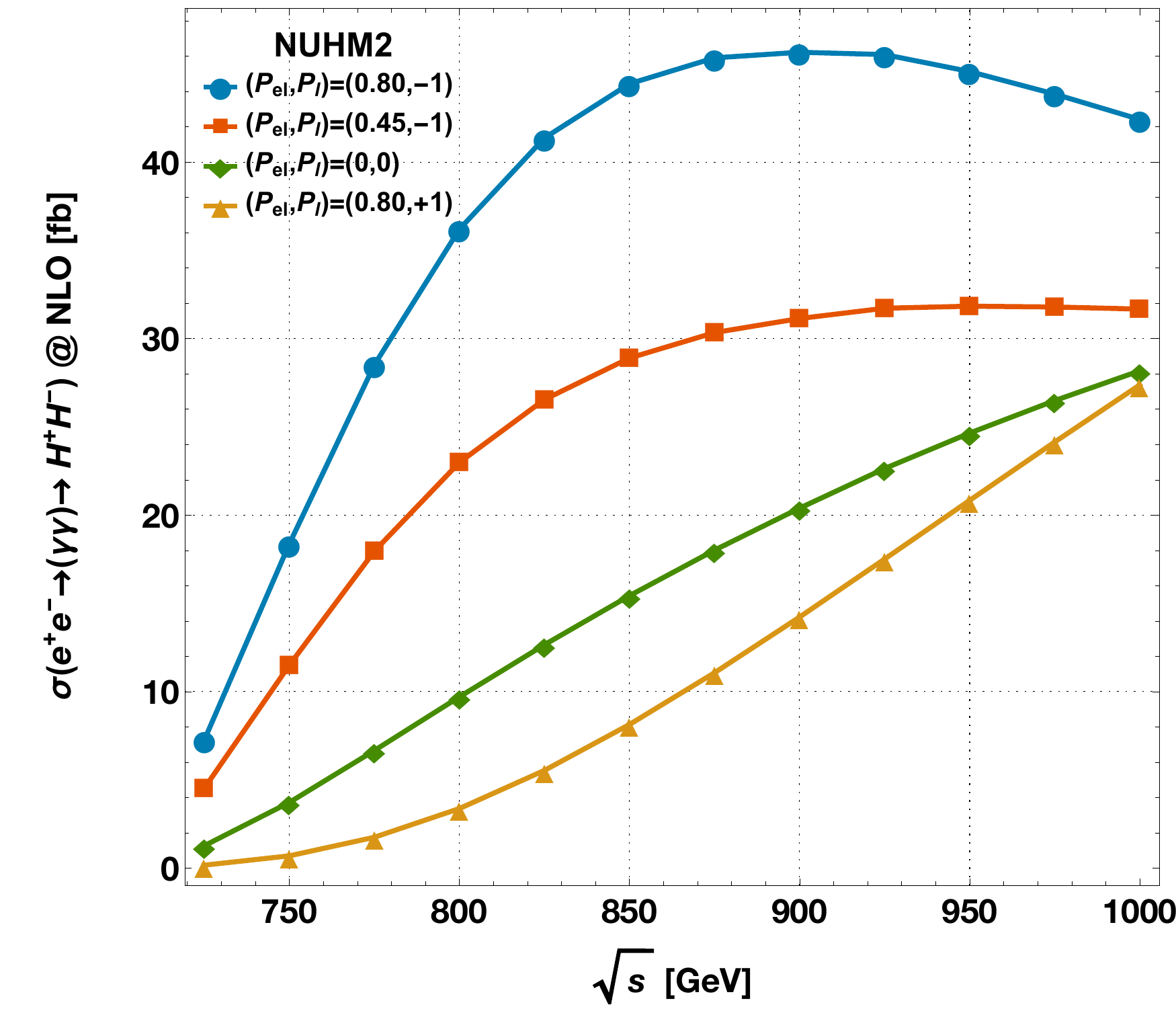}
\caption{\label{fig11} The total cross-section of the charged Higgs pair production convoluted with the photon luminosity in an $e^+e^-$-collider. The polarizations are depicted in the figure.}
\end{figure}
In photon colliders, the main outstanding advantage, which could be underlined, is the possibility of obtaining a high degree of photon polarization. In this study, we calculated the convoluted cross-section assuming various polarizations of the laser beam, $P_\text{l}=(0,\pm1)$, and the electron beam $P_\text{el}=(0,\pm0.45,\pm0.80)$. 
These configurations change the photon energy spectrum of the Compton back-scattered photons, so does the photon luminosity. In Fig. \ref{fig11}, the convoluted cross-section, $\sigma(e^+e^-\rightarrow\gamma\gamma\rightarrow H^+H^)$, is plotted for different polarization configurations of an $e^+e^-$-collider. The convoluted cross-section rises at $\sqrt{s}\gtrsim 720 \gev$ for the benchmark point 1 in all the laser and electron polarization configurations. The rise is dramatic for $(P_\text{el},P_\text{l})=(0.80,-1)$, and it reaches up to $46\fb$ at $\sqrt{s}=0.9\tev$, then it saturates and slowly decreases. Having the opposite polarizations of laser and electron beam increases the number of the high energy photons significantly. 
Comparing the convoluted cross-sections for $(P_\text{el},P_\text{l})=(0,0)$ with the $(P_\text{el},P_\text{l})=(0.80,-1)$ configurations shows that the cross-section is raised by 50\% at $\sqrt{s}=1\tev$.

\begin{figure}[htbp]
\centering
\includegraphics[width=\factor\textwidth]{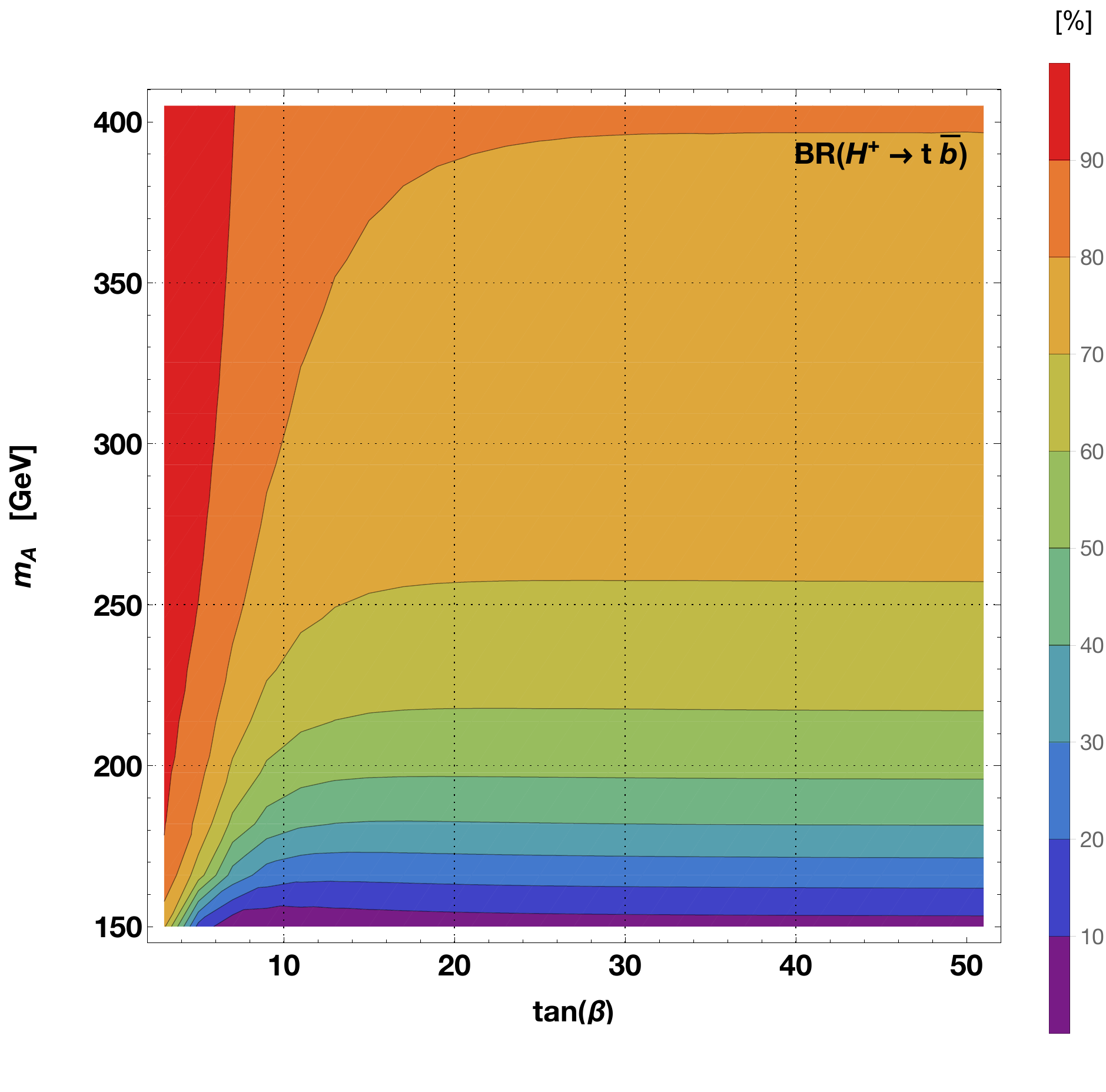}     
\includegraphics[width=\factor\textwidth]{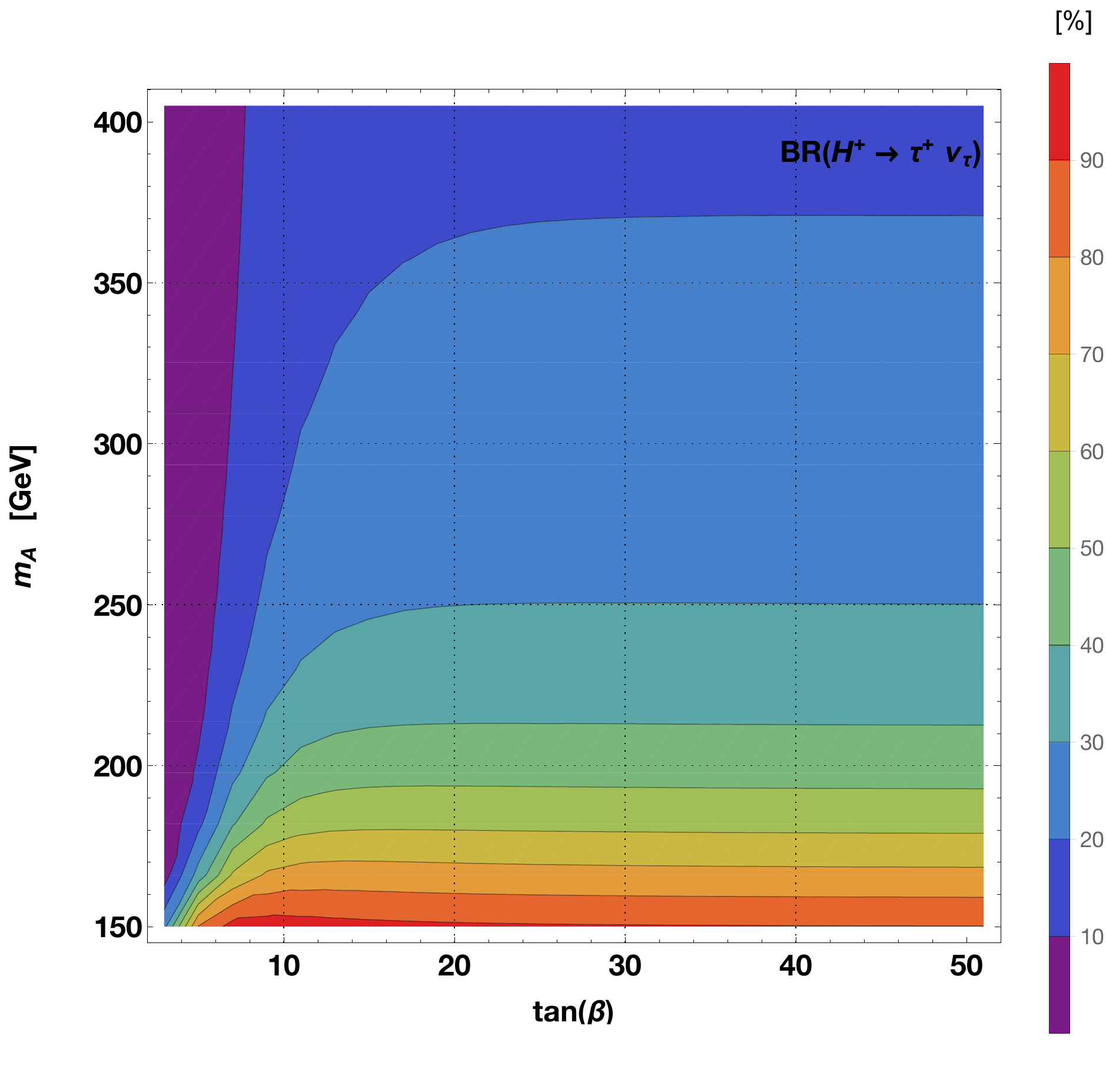}     
\caption{The branching ratio distributions of the charged Higgs boson as a function of $\tb$ and $m_{A^0}$. 
(top): the $H^+\rightarrow  t + \bar{b}$  channel.
(bottom): the $H^+\rightarrow \tau^+ + {\nu}_\tau$ channel.
\label{fig12}
}
\end{figure}
At last, the question whether this signal can be detected at the future colliders need to be answered. The branching ratios of the charged Higgs boson is calculated using \textsc{IsaSugra}. It mainly decays through $H^+\rightarrow  t + \bar{b}$ and $H^+\rightarrow \tau^+ + \nu_\tau$ channels. Their sum gets the lowest 94\% of the total decay width. The distributions are plotted as a function of $\tb$ and $m_{A^0}$ in Fig. \ref{fig12}. At low $m_{A^0}$ values, the channel $H^+\rightarrow \tau^+ + \nu_\tau$ becomes stronger, and $BR(H^+\rightarrow \tau^+ + \nu_\tau)$ gets up to $ 80\%$. If the charged Higgs boson follows this decay channel, then tau could decay hadronically by $\sim64\%$ and leptonically $\sim36\%$ (electron or muon).  The final products of each charged Higgs boson will be tau-jet or $e/\mu$ and missing energy due to neutrinos at the final state. 
The decay channel $H^+\rightarrow  t + \bar{b}$ is picked at  higher rates at high $m_{A^0} \gtrsim 200\gev$ values. 
Since top quark decays through bottom quark and W boson, the final state of the charged Higgs pair will be mainly four b-quark initiated jets and the decay products of the two W bosons.
 

\section{Conclusion and Summary}
\label{sec5}

In this study, the charged Higgs pair production in a $\gamma\gamma$-collider was investigated, including all the virtual one loop and the radiative corrections in the two-parameter non-universal Higgs model. Since the parameters $\tan\beta$ and $m_{A^0}$ are the effective ones among the other parameters of the 2NUHM for getting the masses and the couplings of the Higgses, the results were presented by varying them. 
According to the numerics, the total cross-section increased to $90\fb$.  The total NLO correction is negative overall, and it went down to $-30\%$ in a small region in $\tan\beta$ and $m_A$ plane. The contributions coming from the $\hat{\sigma}_\text{RR} (J_z=0)$ and the $\hat{\sigma}_\text{RL}(J_z=2)$ were calculated, and it was shown that the $J_z=0$ dominates in $\sqrtHatS<0.95\tev$. The total cross-section was convoluted with the photon luminosity in an $e^+e^-$-collider, and the cross-section, $\sigma(e^+e^-\rightarrow\gamma\gamma\rightarrow H^+H^ -)$ reached up to $\sim46\fb$ with $(P_\text{el},P_\text{l})=(0.80,-1)$. Also, comparing the cross-section distributions of different polarization configurations of the laser photon and incoming electron showed that the enhancement rises significantly for the opposite polarizations at every c.m. energy. The results manifest that the charged Higgs pair production in the NUHM2 scenario could be accessed at the ILC with $\sqrt{s}=1\tev$. The charged Higgs boson has two decay channels that are significant for an observation. They are $H^+\rightarrow  t +\bar{b}$ and $H^+\rightarrow \tau^+ + \nu_\tau$ channels. Considering the cross-section of the process and these decay channels, the charged Higgs boson could be detected easily in the future $\gamma\gamma$-collider. The final state consists of a multi-jet environment, where b tagging will play a significant role in studying this process. 
The $\gamma\gamma$-collider hosted on ILC with a minimal cost can provide new insights and distinct mechanism to test the predictions of the supersymmetry, or luckily help to solve the mysteries of the universe.

\section*{Acknowledgements}
The computation presented in this paper is partially performed at TUBITAK ULAKBIM, High Performance and Grid Computing Center (\textsc{TRUBA} resources). Ege University supports this work, project number 17-FEN-054.


\bibliography{template-8s_revtex}

\end{document}